\documentclass[12pt]{iopart}

\usepackage{graphicx}
\usepackage{amssymb}
\usepackage{amsfonts}
\usepackage{bm}
\usepackage{cite}

\begin{document}
\jl{2}

%%%%%%%%%%% my commands: %%%%%%%%%%%%%%%%%%%%
\renewcommand {\i} {{\rm i}}
\renewcommand {\d} {{\rm d}}

\newcommand {\balpha} {\boldsymbol{\alpha}}
\newcommand {\bbeta}  {\boldsymbol{ \beta}}
\newcommand {\bgamma} {\boldsymbol{ \gamma}}
\newcommand {\bnabla} {\boldsymbol{ \nabla}}

\newcommand {\E} {\varepsilon}
\newcommand {\om} {\omega}
\newcommand {\Om} {\Omega}
\newcommand {\cM} {{\cal M}}
\newcommand {\cG} {{\cal G}}
\newcommand {\ee} {{\rm e}}

\newcommand {\bfa} {{\bf a}}
\newcommand {\bfb} {{\bf b}}
\newcommand {\bfe} {{\bf e}}
\newcommand {\bfk} {{\bf k}}
\newcommand {\bfn} {{\bf n}}
\newcommand {\bfp} {{\bf p}}
\newcommand {\bfq} {{\bf q}}
\newcommand {\bfr} {{\bf r}}
\newcommand {\bfs} {{\bf s}}
\newcommand {\bfv} {{\bf v}}
\newcommand {\ra} {{\rm a}}
\newcommand {\rb} {{\rm b}}

\newcommand {\bfD} {{\bf D}}
\newcommand {\bfI} {{\bf I}}

\newcommand {\Pa} {\bfp_{\ra}}
\newcommand {\Pb} {\bfp_{\rb}}
\newcommand {\hc} {\dagger}

\newcommand {\nua} {\nu_{\rm a}}
\newcommand {\nub} {\nu_{\rm b}}

\newcommand {\mua} {\mu_{\rm a}}
\newcommand {\mub} {\mu_{\rm b}}
\newcommand {\xia} {\xi_{\rm a}}
\newcommand {\xib} {\xi_{\rm b}}
\newcommand {\Sa} {\sigma_{\rm a}}
\newcommand {\Sb} {\sigma_{\rm b}}
\newcommand {\tnu} {\tilde{\nu}}
%%%%%%%%%%%%%%%%%%%%%%%%%%%%%%%%%

%%%%%%%%% Title + authors
\title{Relativistic two-photon bremsstrahlung}

\author{A.~V. Korol, I.~A. Solovjev}

\address{Department of Physics,
St.Petersburg State Maritime Technical University,
Leninskii prospect 101, St. Petersburg 198262, Russia}

%%%%%%%%%%%%%%% abstract
\begin{abstract}
An approximate approach for the description of the two-photon bremsstrahlung
emitted by a relativistic projectile scattered in a spherically-symmetric
field is developed.
It based on the accurate treatment of the analytical structure of
the singularities in relativistic one-photon free-free matrix elements.
For the case of a Coulomb field the analytical expressions for the amplitude
and cross section of process are presented.
Numerical results obtained within the framework of the proposed approach
are compared with the available experimental data and with the results
of simpler theories.
\end{abstract}


%\maketitle

%%%%%%%%%%%%%%%%%%%%%%%%%%%%%%%%%%%%%%%%%%%%%%%%
\section{Introduction}
\label{introduction}
%%%%%%%%%%%%%%%%%%%%%%%%%%%%%%%%%%%%%%%%%%%%%%%%

In this paper new results from the relativistic theory of
two-photon bremsstrahlung (2BrS) of a projectile electron
scattered in a spherically-symmetric static field are reported.
We formulate the approximation suitable for effective analytical and
numerical treatment of a rather complex  relativistic free-free
two-photon matrix element.
The developed formalism is applied to the 2BrS process in a point-Coulomb
field.
Results of numerical calculation of the 2BrS cross section
for several geometries of the emission, incident electron energies
and photon frequencies are presented.

The approach, which is described in this paper, based
on the use of a so-called `delta'-approximation.
This method, initially formulated for a non-relativistic
2BrS~\cite{Korol1994a}--\cite{Korol1997} and extended later
to  the case of relativistic collisions~\cite{FedorovaKorolSolovjev2000},
is based on the assumption that to a great extent the behaviour of the
free-free two-photon amplitude is defined by the contributions of
the delta-singular parts of the one-photon  matrix elements
from which the 2BrS amplitude is constructed.

Transitions of such type are encountered not only in the 2BrS problem
but in a number of other physical processes in which the collision
of a projectile with a target  is accompanied
by emission/absorption of photons (see, e.g.,
reviews
\cite{MaquetVeniardMarian1998,EhlotzkyJaronKaminski1998,KorolSolovyov1997}).
The list of such processes includes, in particular,
(a)~the bremsstrahlung-type phenomena in an external field (many-photon
spontaneous bremsstrahlung, laser-induced bremsstrahlung),
(b)~various inelastic processes, when the emission/absorption
of the photon is accompanied by simultaneous excitation or ionization
of the target,
(c)~Compton scattering from many-electron atoms,
(d)~many-photon ionization of atoms and ions.

In the case when perturbation theory in photon-projectile (or
photon-atom) interaction is used to analyze the
above-mentioned processes, the corresponding amplitude ${\cal M}$
can be represented in terms of the compound matrix element
which contains the radiative free-free matrix element between the
intermediate virtual states and in which the integration over
the intermediate momentum is carried out.
The general form of such matrix element is
\begin{equation}
{\cal M}
= \sum_{\mub}
\int \d \Pb\,
{\cal R}_{\mub}(\Pb)\,
D^{(\pm\pm)}_{\nub \nua}(\bfk,\bfe) \, .
\label{3}
\end{equation}
Here $D^{(\pm\pm)}_{\nub \nua}(\bfk,\bfe)$
stands for the matrix element of the one-photon free-free
transition between two relativistic states of a continuous spectrum
with momenta $\Pa$ and $\Pb$ and polarizations
$\mua$ and $\mub$:
\begin{equation}
D^{(\pm\pm)}_{\nub \nua}(\bfk,\bfe)
=
\int \d \bfr\,
\Psi^{(\pm)\,\hc}_{\nub}(\bfr)\,
(\bfe\balpha)\, \exp{(-\i \bfk\bfr)}
\, \Psi^{(\pm)}_{\nua}(\bfr)\,.
\label{2}
\end{equation}
Here $\Psi^{(\pm)}_{\nu}(\bfr)$ are the bispinor wavefunctions
corresponding to the out- (the upper index `$+$') and to the in
(`$-$') scattering states,
the symbol $\hc$  denotes the hermitian conjugation,
$\balpha = \gamma^0\bgamma$ where $\gamma^0$, $\bgamma$ are
the Dirac matrices.
The vectors $\hbar\bfk$ and $\bfe$ denote the photon momentum and
polarization.
We choose the gauge in which $\bfe\bfk=0$.
The CGS system is used throughout the paper.

The factor ${\cal R}_{\mub}(\Pb)$ represents the con\-tri\-bu\-tion of all
the remaining processes related to the collision.
In particular, in the case of 2BrS process,
${\cal R}_{\mu_{\rm b}}(\Pb)$ contains another
free-free matrix element (see general expression~(\ref{1.1a})
for the 2BrS amplitude).

Relation (\ref{2}) defines four matrix elements the type of which
depends on the asymptotic behaviour of the wavefunctions
of the initial and final states.
In the vicinity of the points defined by the relations
\begin{eqnarray}
\Pa - \Pb - \hbar\bfk = 0 \, ,
\ \quad
p_{\ra} - \left|\Pb + \hbar\bfk\right| = 0 \, ,
\ \quad
|\Pa - \hbar\bfk| - p_{\rb} = 0 \ ,
\label{c1}
\end{eqnarray}
the matrix elements ${D^{(\pm\pm)}_{\nub \nua}(\bfk,\bfe)}$ contain
the singular terms
of the two essentially different types~\cite{FedorovaKorolSolovjev2000}.
Firstly, there are well known pole-like singularities which are
responsible, in particular, for the infrared divergency of
the one-photon bremsstrahlung amplitude (see, e.g.,
\cite{Land4,Akhiezer,Rosenberg1994}.
In this case both momenta, $\Pa$ and $\Pb$, lie on
the mass surface, so that the exact equalities (\ref{c1}) are inconsistent
with the energy conservation law, which can be written as
$\E_{\rm a}=\E_{\rm b} + \hbar\,\om$
(here $\E_{\rm a,b}$ are the energies of the projectile in the states
'a' and 'b', $\hbar\,\om$ is the photon energy).
Hence, in this case the equality signs must be understood as the limits
$\to 0$.

The singularities of the second type
(we call them  `delta' singularities)
contain terms proportional to the delta-functions
$\delta(\Pa - \Pb  -  \hbar\bfk)$,
$\delta(p_{\ra}  -  |\Pb  + \hbar\bfk |)$ and
$\delta(|\Pa  - \hbar\bfk|  -  p_{\rb})$
(note, that the first of these is the three-dimensional
$\delta$-function, whereas the two others are one-dimensional).
In the compound matrix elements (\ref{3}) one of the states `a' or `b'
(or both, if the process includes the emission of two and more photons, or
if the photon emission occurs between two virtual states)
describes the off-mass surface particle.
Therefore, the `delta' terms add a non-zero contribution to the amplitude
(\ref{3}).

The approach for the approximate treatment
(the `delta'-approximation) of the compound matrix elements of the
type (\ref{3}) is based on the assumption that to a great extent
the behaviour of ${\cal M}$ is defined by the contributions of
the delta-singular parts of the one-photon free-free matrix elements
$D^{(\pm\pm)}_{\nub \nua}(\bfk,\bfe)$.
Initially~\cite{Korol1994a,Korol1995a,Korol1996}
it was formulated in connection with the problem of two-photon dipole
free-free transitions  of a non-relativistic projectile moving in
an external field of atomic target.
In the cited papers as well as in~\cite{Korol1997,Korol1995b}
it was demonstrated that over wide regions of the incident electron energies
and the photon frequencies the approximate approach produces nearly the
same results as the rigorous treatment for both dipole
\cite{FlorescuDjamo1986,VeniardGavrilaMaquet1987,GavrilaMaquetVeniard1990,
DonderaFlorescu1993}
and non-dipole~\cite{Korol1997,DonderaFlorescu1998}  photons.

Within the framework of the non-relativistic dipole-photon
approximation it was demonstrated
that for the correct evaluation of (\ref{3}) it is necessary to take
into account all singular terms of the free-free matrix elements, and
in many cases the contribution of the delta-terms to the integral on
the right-hand side is highly noticeable, if not dominant. Such an analysis
has been carried out in connection with many-photon ionization
(detachment)
\cite{VeniardPiraux1990,BlahaDavis1995,MercourisKomninosDionissopoulouNicolaides1996,GribakinIvanovKorolKuchiev1999_2000},
the single-photon bremsstrahlung process with simultaneous ionization
of the target~\cite{Korol1994c}, the process of two-photon bremsstrahlung
\cite{ManakovMarmoFainstein1995,KrylovetskiiManakovMarmoStarace2002},
and the process of
spontaneous bremsstrahlung in presence of a laser field
\cite{FlorescuFlorescu2000}.
In non-relativistic  non-dipole approximation the analogous treatment
has been applied so far to the free-free compound matrix element which
characterizes the amplitude of two-photon bremsstrahlung
\cite{Korol1996,Korol1997}. 

A detailed review of the achievements of 
the non-relativistic theory of 2BrS
can be found in~\cite{Korol1997,FedorovaKorolSolovjev2000}.
Here we just mention the main results obtained in this field.

Theoretical activity was stimulated by a
series of experiments 
\cite{AltmanQuarles1985,LehtihetQuarles1989,Kahler92,Quarles93a,Quarles93b,
Quarles93c,Hippler1991,HipplerSchneider1994}
 in which the data on the 2BrS cross
section and angular distribution were obtained for relatively high
energies of the incident electron and emitted photon (up to 70 keV).
Apart from the early work of Smirnov \cite{Smirnov1977}, 
where the 2BrS process
was treated in terms of the relativistic first Born
approximation (see also \cite{GhilenceaToaderDiaconu1995}), 
theoretical investigations were based on second order
non-relativistic perturbation theory.
Within the framework of this approximation the exact analytical
formulae were obtained for the compound free-free matrix element in
the Coulomb field.
It was done in the dipole-photon case 
\cite{GavrilaMaquetVeniard1985,GavrilaMaquetVeniard1990,FlorescuDjamo1986,
VeniardGavrilaMaquet1987,ManakovMarmoFainstein1995}
and with the retardation effects included
\cite{Korol1997,DonderaFlorescu1998} as well.
For neutral atomic targets the calculations of the 2BrS spectrum were
performed within the frame of the potential (`ordinary')
bremsstrahlung model \cite{Korol1993a,Korol1994a,Korol1995b}
as well as with the 2BrS emission due to the
`polarizational' bremsstrahlung mechanism 
 taken into account \cite{AmusiaKorol1993,KrackeEtAl1993,KrackeEtAl1994}.

In connection with the experiments 
\cite{Kahler92,Quarles93a,Quarles93b,Quarles93c}, where the 2BrS process was
studied for 70 keV electrons scattered by various many-electron atoms,
it was noted in several publications
\cite{Korol1994a,Korol1996,Korol1997,DonderaFlorescu1993,DonderaFlorescu1998,
DonderaFlorescuPratt1996}
that the role of the retardation effects in the 2BrS process is much
higher than in the conventional one-photon bremsstrahlung.
In these papers it was also demonstrated that 
that for the energies of the electron and the photons within the
tens of keV range (i.e. as in the cited experiments)
it is the retardation and relativistic effects which are mostly important
for the formation of the 2BrS spectrum
rather than the effect of screening due to atomic electrons.
The inclusion of the latter effect in the calculation scheme
mofifies the result on the level of several per cent (see, e.g.,
\cite{Korol1994a}) whereas the account for the radiation retardation 
increases the 2BrS cross section by the order of magnitude
\cite{Korol1997,DonderaFlorescu1998}.
Therefore, to obtain a reliable theoretical result one has not only go
beyond the frame of the non-relativistic dipole-photon approximation
but  to consider the fully relativistic treatment of the problem.

The relativistic treatment of the compound
many-photon transitions is much more complicated than its non-relativistic
analogue from both analytical and computational viewpoints
\cite{Gorshkov1964b,FlorescuGavrila1976,WongYeh1985,
ZapryagaevManakovPalchikov1985,ManakovOvsyannikovRappoport1986,
MuCrasemann1988,DonderaMarineskuPratt1990,
BergstromSuricPiskPratt1993,ManakovMaquetMarmoSzymanovski1998,
Szmytkowski2002,Yakhontov2003,
Szmytkowski2004}.
In the cited papers the theory and the numerical results for various
processes described by the two-photon bound-bound and bound-free transitions
are presented.
In contrast, the exact relativistic treatment of the free-free two-photon
transitions has not been suggested so far.
Till recently all theoretical considerations did not go beyond the framework
of the plane wave first Born
approximation~\cite{Smirnov1977,GhilenceaToaderDiaconu1995} and
the soft-photon limit~\cite{Rosenberg1987_1990_1991}.
In the recent paper~\cite{FedorovaKorolSolovjev2000} the relativistic formalism
of `delta'-approximation for treatment of many-photon free-free transitions
was developed. 
Its application was illustrated by the evaluation of the delta-amplitude
of 2BrS process. 
Excluding the papers
\cite{Smirnov1977,GhilenceaToaderDiaconu1995,Rosenberg1987_1990_1991}
there were no numerical results of relativistic calculations of 2BrS cross
sections.

%In what follows
In order to fill this gap we extend the approximate treatment
of the relativistic 2BrS
developed in~\cite{FedorovaKorolSolovjev2000} to the case of the point
Coulomb field, $-Ze/r$, and perform a numerical analysis of the role of 
retardation and relativistic effects in the formation of spectral and
angular distributions of 2BrS for the conditions of the 
experiments~\cite{Kahler92,Quarles93a,Quarles93b,Quarles93c}.
This is done within the framework of the `delta'-approximation, described
in more detail in~\cite{FedorovaKorolSolovjev2000} and in
section~\ref{formalism} below.
Because of the absence of the closed analytic expression for the relativistic
scattering states wavefunctions we used the Furry-Sommerfeld-Maue (FSM)
wavefunctions~(e.g.~\cite{Land4}). The applicability of FSM approximation
in view of available experimental data is discussed in section~\ref{FSM}.

%In spite of the fact that the validity of the `delta'-approximation
%has been not proved formally,
%the successful treatment of the above-mentioned processes within the
%framework of this approach can be considered as an indirect confirmation
%of the adequacy of the `delta'-approximation and the relevancy of its use
%alongside with other widely used approximations.
%%Besides this
%In the particular case of 2BrS process it was also demonstrated
%(for both non-relativistic~\cite{Korol1996} and
%relativistic~\cite{FedorovaKorolSolovjev2000} 2BrS) that the general
%expression for delta-amplitude of the process correctly reproduces the
%important limiting cases: the plane-wave first Born approximation and
%the soft-photon approximation.

In spite of the fact that the validity of the `delta'-approximation
has  not been formally proved there is a number of indirect confirmations of
the applicability of this approach. 
In particular it was demonstrated
(both in the  non-relativistic~\cite{Korol1996} and
relativistic~\cite{FedorovaKorolSolovjev2000} cases) that the general
expression for the 'delta'-amplitude of the 2BrS process correctly 
reproduces the important limiting cases: 
the plane-wave first Born approximation and the soft-photon
approximation. 
Also the `delta'-approximation, being applied for the description
of a number of mentioned above processes, yields reasonable numerical
results against the results of the exact calculations.
In the dipole-photon regime it was done for the two-photon bremsstrahlung
\cite{Korol1994a,Korol1995a,Korol1995b,KrylovetskiiManakovMarmoStarace2002}, 
two- and three-photon detachment
of electrons from negative ions~\cite{GribakinIvanovKorolKuchiev1999}.
Beyond
the dipole-photon approximation the comparison was carried out for the 2BrS
process of a non-relativistic electron~\cite{Korol1996,Korol1997}.

This approach, although being approximate, allows us to evaluate effectively
the principal parts of the free-free two-photon matrix element in the
relativistic domain with much less analytical and computational efforts.
The method  can be generalized to embrace the $n$-photon ($n>2$) radiative
free-free transitions.

The formalism is described in section~\ref{formalism}.
In section~\ref{results} we present the results of numerical calculations
and carry out the comparison with available experimental data and
the results of simpler theories.

%%%%%%%%%%%%%%%%%%%%%%%%%%%%%%%%%%%%%%%%%%%%%%%%
\section{Formalism}
\label{formalism}
%%%%%%%%%%%%%%%%%%%%%%%%%%%%%%%%%%%%%%%%%%%%%%%%

%%%%%%%%%%%%%%%%%%%%%%%%%%%%%%%%%%%%%%%%%%%%%%%%
\subsection{2BrS amplitude and cross section: general formulae}
\label{gen_expr}
%%%%%%%%%%%%%%%%%%%%%%%%%%%%%%%%%%%%%%%%%%%%%%%%

%%%%%%%%%%%%%%%%%%%%%%%%%%%%%%%%%%%%%%%%%%%%%%%%%%%%%%%%%%%%%%%%%%%%%%%%%%%
\begin{figure}
\begin{center}
\includegraphics[width=10cm,height=2cm,angle=0]{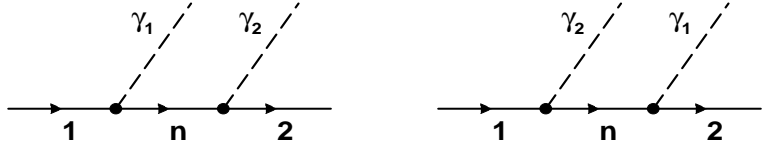}
\end{center}
\caption{Diagrammatical representation of the 2BrS process.
The solid lines correspond to the relativistic projectile
moving in an external  field of the target (the Furry picture,
e.g.~\cite{Land4}).
The transition from  the initial state  `1' into the final
state `2'  via the intermediate (virtual) state `n'
is accompanied by the emission of two photons $\gamma_{1,2}$
(the dashed lines).}
\label{diag_2BrS}
\end{figure}
%%%%%%%%%%%%%%%%%%%%%%%%%%%%%%%%%%%%%%%%%%%%%%%%%%%%%%%%%%%%%%%%%%%%%%%%%%%

The 2BrS process during the
potential scattering of an electron (the charge is labeled as $e$ and
the mass as $m_{\rm p}$) is a transition of the projectile from
the initial state $\nu_1=(\bfp_1,\mu_1)$
to the final state $\nu_2=(\bfp_2,\mu_2)$  accompanied by
the emission of two photons.
Each of the photons  ($j=1,2$) is characterized by the energy,
$\hbar\om_j$, momentum $\hbar\bfk_j$ and the polarizational vector $\bfe_j$.
The energies of all particles satisfy the energy conservation law
\begin{equation}
\E_1 = \E_2 + \hbar\om_1 + \hbar\om_2 \ .
\label{cons_law_2BrS}
\end{equation}
where $\E_{1,2} = \sqrt{p_{1,2}^2 c^2 + m_{\rm p}^2c^4}$ are the
particle energies in the initial and the final states.

The four-fold cross section differential with respect to
$\om_{1,2}$ and to the solid angles of the photons emission
$\Om_{1,2}$
is given by
\begin{equation}
{\d^4\sigma \over \d\om_1\,\d\om_2\, \d\Om_1\,\d\Om_2}
=
{\alpha^2 \over 2}\,
{\E_1\E_2 \over c^4}\,
{p_2 \over p_1}\,
{\om_1\, \om_2 \over (2\pi)^6}\,
\sum_{\lambda_1,\lambda_2}
\sum_{\mu_1,\mu_2}
\int_{(4\pi)} \d\Om_{\bfp_2}\,
\left|\cM\right|^2
\, .
\label{CS_2BrS}
\end{equation}
Here $\alpha$ is the fine structure constant,
the integration is performed over the solid angle of the scattered
particle,
the summations are carried out over the polarizations of the photons
($\lambda_{1,2}$) and of the projectile ($\mu_{1,2}$).

The total amplitude of the process, $\cM$, is described by
two Feynman diagrams as presented in figure~\ref{diag_2BrS}.
Each diagram corresponds to the compound matrix element of
the two-photon transition from the initial state `1' via the
intermediate (virtual) state `$n$' to the final state `2':
\begin{eqnarray}
\fl
\cM = \cM_{[12]} + \cM_{[21]} \, ,
\quad
\cM_{[12]}
=
\sum_n
{
\langle \bfp_2 \mu_2 |
\bfe_2\bgamma\, \ee^{-\i \bfk_2 \bfr}
|n\rangle
\langle n|
\bfe_1\bgamma\, \ee^{-\i \bfk_1 \bfr}
| \bfp_1\mu_1\rangle
\over
\E_n + \hbar\om_1 - \E_1 - \i 0
} \, . \quad
\label{1.1a}
\end{eqnarray}
Here $\langle \bfp_2 \mu_2|
\equiv \Psi_{\bfp_2\mu_2}^{(-)\,\hc}(\bfr)\,\gamma^{0}$ and
$|\bfp_1\mu_1\rangle\equiv \Psi_{\bfp_1\mu_1}^{(+)}(\bfr)$
are the projectile's initial- and final-state wavefunctions.
The sum is carried out over the complete set of states
$|n\rangle$ of the Hamiltonian
$\hat{H} = c \balpha\hat{\bfp} + \gamma^0 mc^2 + V(r)$,
and includes the contributions of the positive-energy
($\E_n>0$) and the negative-energy ($\E_n<0$) states,
the index $n$ includes all quantum numbers which characterize
the intermediate state.
The term $\cM_{[21]}$  can be obtained from $\cM_{[12]}$
(the ordering of the subscripts inside the square brackets indicates
which of the photons was emitted first) by
exchanging $(\om_1,\bfk_1,\bfe_1)\leftrightarrow (\om_2,\bfk_2,\bfe_2)$.

The bispinor wavefunctions ${\Psi_{\bfp\mu}^{(\pm)}(\bfr)}$, which
describe the scattering states with the asymptotic momenta $\bfp$,
can be expanded in the partial-wave series over the functions
$\Psi_{\E j l m}(\bfr)$ characterized by energy, $\E$, total
angular momentum, $j$, orbital momentum, $l$, and projection of
the total momentum, $m$~\cite{Akhiezer}:
\begin{eqnarray}
\Psi_{\bfp\mu}^{(\pm)}(\bfr)
=
{4\pi\hbar \over p}\,
\sum_{j l m}
\left(\Om_{j l m}^{\hc}(\bfn_{\bfp})\, v_{\mu}(\bfn_{\bfp})\right)
\ee^{\pm \i\delta_{j l}(\E)}\,
\Psi_{\E j l m}(\bfr)\, .
\label{ord1.4a}
\end{eqnarray}
Here the notation $\bfn_{\bfa}$ stands for the unit vector along
the direction $\bfa$,
$\Om_{j l m}(\bfn)$ denotes a spherical spinor defined as in~\cite{BMX},
$v_{\mu}$ is a unit two-component spinor corresponding to the spin
projection $\mu$.
The quantities $\delta_{jl}(\E)$ are the relativistic scattering
phaseshifts.

The non-relativistic analogue of ${M_{[12]}}$
was investigated analytically (in the case of a point Coulomb field)
\cite{Korol1997,FlorescuDjamo1986,VeniardGavrilaMaquet1987,
GavrilaMaquetVeniard1990,DonderaFlorescu1998,
ManakovMarmoFainstein1995,KrylovetskiiManakovMarmoStarace2002,
GavrilaMaquetVeniard1985}
and numerically
(the cited papers and also
\cite{Korol1994a,Korol1995b,Geltman1996,Korol1993a}).

Contrary to the non-relativistic case,
the evaluation of the right-hand side of $\cM_{[12]}$ from
(\ref{1.1a}) can be envisaged only by means of numerical calculations.
Even in the case of a point Coulomb field one can
hardly anticipate any real progress in applying analytical methods.
This is due to both the complexity of the analytic structure of the
relativistic Coulomb Green's function (e.g., see the discussion in a
recent review~\cite{MaquetVeniardMarian1998})
and to the absence of the closed analytic expressions for the
scattering states wavefunctions (\ref{ord1.4a}).

The direct numerical computation is also a challenging problem.
So far real progress has been achieved only in the fully relativistic
treatment of the bound-bound and bound-free compound matrix elements (see,
e.g.,
\cite{BergstromSuricPiskPratt1993,ManakovMaquetMarmoSzymanovski1998,
Szmytkowski2002,Yakhontov2003,Szmytkowski2004}
and references therein).
Even though it is possible, using  (\ref{ord1.4a}) and the corresponding
expansions for the Green's function and for the operators
$\bfe_{1,2}\bgamma\, \ee^{-\i \bfk_{1,2} \bfr}$,
to write down the partial wave expansion for the amplitude $\cM$,
it is still a formidable task of computing the obtained multi-fold series.

Alternatively, to compute the exact relativistic
2BrS amplitude one can apply the method due to Sternheimer~\cite{Sternheimer}
and Dalgarno and Lewis~\cite{DalgarnoLewis}.
In this case, instead of the direct evaluation of the sum over the
intermediate state $n$, one can evaluate $\cM_{[12]}$
as a matrix element of the operator
${\bfe_2\bgamma\,\ee^{-\i \bfk_2 \bfr}}$ between the final
state $\langle \bfp_2 \mu_2|$ and some auxiliary function
which is the solution of properly defined inhomogeneous Dirac equation.
In the relativistic domain this method was successfully applied
to the bound-free two-photon transitions
\cite{BergstromSuricPiskPratt1993}, and
to the bound-bound ones~\cite{Yakhontov2003,Szmytkowski2004}.
On the basis of preliminary analytical and numerical work which we have
done (but do not report on it in this paper)
it seems feasible to compute the 2BrS amplitude by this method.

However, the aim of this paper is to apply another approach,
which, although being approximate,  allows one to carry out the analysis
of the characteristics of the relativistic 2BrS with much less
analytical and computational efforts.
For the sake of convenience, in the next section
we briefly outline the idea and the basic formulae of the approximate
method~\cite{FedorovaKorolSolovjev2000} which is used further in the
paper for the calculation  of the 2BrS cross section of
a relativistic electron in a point Coulomb field.

%%%%%%%%%%%%%%%%%%%%%%%%%%%%%%%%%%%%%%%%%%%%%%%%
\subsection{2BrS amplitude in the `delta'-approximation}
\label{delta}
%%%%%%%%%%%%%%%%%%%%%%%%%%%%%%%%%%%%%%%%%%%%%%%%

To introduce the `delta'-approximation let us start with
separating the contributions of the positive- and
negative-energy parts of the electronic propagator
to the total matrix element $\cM_{[12]}$ from (\ref{1.1a}):
\begin{eqnarray}
\cM_{[12]} = \left[\cM_{[12]}\right]_{+} + \left[\cM_{[12]}\right]_{-}
\label{delta.1}
\end{eqnarray}

To avoid the unnecessary complications we assume that
the positive-energy spectrum of the electron in the external field
$V(r)$ does not contain the bound states, so that
all intermediate states with $\E_{n}>0$ belong to the
continuous spectrum.
Hence, the sum on the right-hand side of (\ref{1.1a})
can be understood as the sum over the
bispinor polarizations $\mu$ and the integral over the momenta
${\bfp}$.
Hence, the amplitude $\left[\cM_{[12]}\right]_{+}$ can be written as
follows:
\begin{eqnarray}
\left[\cM_{[12]}\right]_{+}
=
\sum_{\mu}
\int
{\d \bfp \over (2\pi \hbar)^3}\,
{D_{\nu_2 \nu}^{(-\pm)}(\bfk_2,\bfe_2)\,
D_{\nu \nu_1}^{(\pm+)}(\bfk_1,\bfe_1)
\over
\E_{\nu}+\hbar\,\om_1 -\E_1- \i 0
}
=
R_{[12]}
+
\i\,I_{[12]}\, .
\label{delta.2}
\end{eqnarray}
The term $I_{[12]}$
(which appears if one uses the standard rule to extract the
imaginary part of the integrand's denominator)
describes a two-step emission process in which the
energy of the intermediate state satisfies the energy conservation
law: $\E_{\nu} = \E_1 - \hbar\om_1$.
This quantity is given by the following integral
\begin{equation}
I_{[12]}
=
\left[
{p \, \E \over 8 \pi^2 \hbar^3 c^2}\,
\sum_{\mu}
\int
\d \Omega_{\bfp} \,
D_{\nu_2\nu}^{(-\pm)}(\bfk_2,\bfe_2)\,
D_{\nu\nu_1}^{(\pm+)}(\bfk_1,\bfe_1)
\right]_{\E_{\nu}= \E_1 - \hbar \om_1}\, .
\label{delta.4}
\end{equation}
We note that because of the angular integral over $\Omega_{\bfp}$
the result of integration in (\ref{delta.4}), as well as in (\ref{delta.2}),
does not depend on the type of basic set (i.e. `$+$' or `$-$')
chosen to describe the intermediate continuum states $\Psi_{\nu}(\bfr)$.

The approximation, briefly described below (for the details see
\cite{Korol1996,FedorovaKorolSolovjev2000}), concerns
the evaluation of the first term, $R_{[12]}$, from (\ref{delta.2}) which
reads
\begin{equation}
R_{[12]}
=
\sum_{\mu}
{\rm v.p.}
\int
{\d \bfp \over (2\pi \hbar)^3}\,
{
D_{\nu_2\nu}^{(-\pm)}(\bfk_2,\bfe_2)\,
D_{\nu\nu_1}^{(\pm+)}(\bfk_1,\bfe_1)
\over
\E_{\nu}+\hbar\om_1 -\E_1
}\, ,
\label{delta.5}
\end{equation}
where {\rm v.p.} indicates the principal value integration with respect to
the pole coming from the integrand's denominator.
The free-free matrix elements $D_{\nu_2 \nu}^{(-\pm)}(\bfk_2,\bfe_2)$ and
$D_{\nu \nu_1}^{(\pm+)}(\bfk_1,\bfe_1)$ correspond to
the radiative {\it virtual} transition.
The momentum $\bfp$ of the intermediate state $\nu$
is not fixed by any conservation law.
The delta-singular terms in $D_{\nu_2 \nu}^{(-\pm)}(\bfk_2,\bfe_2)$
appear (cf. eq. (\ref{c1})) if $\bfp$ satisfies either
$\left|\bfp_2 + \hbar\bfk_2\right| - p = 0$ or
$\left|\bfp - \hbar\bfk_2\right| - p_2 = 0$, which define two spheres
in the momentum space.
In the vicinity of these spheres the matrix element has pole-like
singularity which must be treated in a principal-value sense
when carrying out the integration over $\bfp$.
The singular properties of $D_{\nu \nu_1}^{(\pm+)}(\bfk_1,\bfe_1)$
are similar but related to another pair of spheres:
$\left|\bfp + \hbar\bfk_1\right| - p_1 = 0$ and
$\left|\bfp_1 - \hbar\bfk_1\right| - p = 0$.
The $\delta$-function terms from both of these matrix elements
add a non-zero contribution to the integral from (\ref{delta.5}).
As the result, one can  express $R_{[12]}$ as a sum of two terms.
The first term, $R_{[12]}^{(\delta)}$ is only due to the contribution
of the $\delta$-singular parts of the matrix elements, whereas
the second one, $R_{[12]}^{{\rm (reg)}}$, accounts for the integration
over the whole $\bfp$-space without the points which belong
to the spheres defined above.

Within the framework of the `delta'-approximation the term
$R_{[12]}^{{\rm (reg)}}$
is omitted completely, so that instead of $R_{[12]}$ one uses only
$R_{[12]}^{(\delta)}$:
\begin{eqnarray}
\fl
R_{[12]}
\longrightarrow
R_{[12]}^{(\delta)}
=
{1\over 2}
\sum_{\mu}
&&
\Biggl\{
\left[
{\bfe_2
\left(\bfD^{(-+)}_{\nu_2 \tnu}(\bfk_2)
+
\bfD^{(--)}_{\nu_2 \tnu}(\bfk_2)
\right)
\over
\E_s + \hbar\om_1 - \E_1}
\, b_{\tnu\nu_1}(\bfe_1)
\right]_{\bfs=\bfp_1-\hbar\bfk_1}
\nonumber\\
&&
+
\left[
b_{\nu_2\tnu}(\bfe_2)
\,
{
\bfe_1
\left(
\bfD^{(-+)}_{\tnu \nu_1}(\bfk_1)
+
\bfD^{(++)}_{\tnu \nu_1}(\bfk_1)
\right)
\over
 \E_s + \hbar\om_1 - \E_1}
\right]_{\bfs=\bfp_2+\hbar\bfk_2}
\nonumber\\
&&
+
{\i p_1 \over 2 \pi \hbar}
\int \d \Om_{\bfp}
\left[
{
\bfe_2  \bfD^{(--)}_{\nu_2 \tnu}(\bfk_2)\,
f^{(+)}_{\tnu \nu_1}(\bfn_{\bfp},\bfe_1)
\over
 \E_s + \hbar\om_1 - \E_1}
 \right]_{\scriptstyle{ \bfs=p_1\bfn_{\bfp}-\hbar\bfk_1 }}
\nonumber\\
&&
+
{\i p_2 \over 2 \pi \hbar}
\int \d \Omega_{\bfp}
\left[
{ f^{(-)\, \hc}_{\tnu \nu_2}(-\bfn_{\bfp},\bfe_2)\,
\bfe_1  \bfD^{(++)}_{\tnu \nu_1}(\bfk_1)
\over
\E_s + \hbar\om_1 - \E_1}
\right]_{\scriptstyle{ \bfs=p_2\bfn_{\bfp}+\hbar\bfk_2 }}
\Biggr\}
\, .
\label{delta.8}
\end{eqnarray}
Here the subscript $\tnu$ stands for the set ${(\bfs,\mu)}$ with
the momentum $\bfs$ defined as indicated.
The vector matrix element
$\bfD^{(\sigma_{\rb}\sigma_{\ra})}_{\nub \nua}(\bfk)$
(where $\sigma_{\rb}=`\pm'$ and $\sigma_{\rb}=`\pm'$)
is given by the integral
\begin{equation}
\bfD^{(\sigma_{\rb}\sigma_{\ra})}_{\nub \nua}(\bfk)
=
\int \d \bfr\,
\Psi^{(\Sb) \hc}_{\nub}(\bfr)\,
\balpha\, \exp{(- \i \bfk \bfr)}\,
\Psi^{(\Sa)}_{\nua}(\bfr)\, ,
\label{all2.2}
\end{equation}
and is subject to the conditions
$\left|\Pb + \hbar\bfk\right| - p_{\ra} \neq 0$ and
$\left|\Pa - \hbar\bfk\right| - p_{\rb} \neq 0$,
which mean that the matrix elements
on the right-hand side of (\ref{delta.8}) do not contain
the $\delta$-terms.
The quantities $f^{(\pm)}_{\nub\nua}(\bfn,\bfe)$
and $b_{\nub\nua}(\bfe)$ are expressed in terms of the unit
bispinor $u_{\nu}$ of a plane wave, and
the bispinor scattering amplitude $\cG^{(\pm)}_{\nu}(\bfn_{\bfr})$
\cite{Akhiezer}:
\begin{eqnarray}
f^{(\pm)}_{\nub\nua}(\bfn,\bfe)
=
u^{\hc}_{\nub}\,(\bfe\balpha)\,
\cG^{(\pm)}_{ \nua}(\bfn)
\, ,
\qquad
b_{\nub\nua}(\bfe)
=
u^{\hc}_{\nub}\, (\bfe\balpha)\, u_{\nua} \, .
\label{A13c}
\end{eqnarray}

Accounting for (\ref{delta.8}) one substitutes the exact
amplitude $\left[\cM_{[12]}\right]_{+}$ with the quantity
$\left[\cM_{[12]}^{(\delta)}\right]_{+}$ according to the rule:
\begin{eqnarray}
\left[\cM_{[12]}\right]_{+}
\longrightarrow
\left[\cM_{[12]}^{(\delta)}\right]_{+}
=
R_{[12]}^{(\delta)}
+
\i\,I_{[12]}\, ,
\label{delta.2a}
\end{eqnarray}
with $I_{[12]}$ defined in (\ref{delta.4}).

Hence, within the framework of the `delta'-approximation
the total 2BrS amplitude $\cM$ (see (\ref{1.1a}))
acquires the form:
\begin{eqnarray}
\cM
\longrightarrow
\cM ^{(\delta)}
=
R_{[12]}^{(\delta)}
+
R_{[21]}^{(\delta)}
+
\i\Bigl(I_{[12]}+I_{[21]}\Bigr)
+
\left[\cM_{[12]}\right]_{-}
+
\left[\cM_{[21]}\right]_{-} \, .
\label{delta.12}
\end{eqnarray}
Here one can account for the approximate formula derived in
\cite{FedorovaKorolSolovjev2000}:
\begin{eqnarray}
\left[{\cal M}_{[12]}\right]_{-}
+
\left[{\cal M}_{[21]}\right]_{-}
\simeq
-{\bfe_1 \bfe_2 \over m_{\rm p} c^2}\,
\int
\d \bfr\,
 \Psi^{(-)\, \hc}_{\nu_2}(\bfr)\,
\ee^{-\i(\bfk_1+\bfk_2)\bfr}\,
\Psi^{(+)}_{\nu_1}(\bfr)\,.
\label{sg.6}
\end{eqnarray}

%%%%%%%%%%%%%%%%%%%%%%%%%%%%%%%%%%%%%%%%%%%%%%%%%%%%%%%%%%%%%%%%%%%%%%%%%
\subsection{Application to a point Coulomb field}
\label{FSM}
%%%%%%%%%%%%%%%%%%%%%%%%%%%%%%%%%%%%%%%%%%%%%%%%%%%%%%%%%%%%%%%%%%%%%%%%%

In this section we apply the developed approach to construct the
approximate amplitude of the 2BrS process in the case of a
relativistic electron scattering in a point Coulomb field, $-Ze/r$.
This case is of interest in connection with the experimental data
obtained for $\E_1 = 70$ keV electrons scattered
by various targets~\cite{Kahler92,Quarles93a,Quarles93b,Quarles93c}.
It was  noted in several publications
\cite{Korol1994a,Korol1996,Korol1997,DonderaFlorescu1993,DonderaFlorescu1998,
DonderaFlorescuPratt1996}
that for the energies of the electron and the photons within the
tens of keV range (i.e. as in the experiments)
it is the retardation and relativistic effects which are mostly important
for the formation of the 2BrS spectrum
rather than the effect of screening due to atomic electrons.

The additional difficulty (as compared to the non-relativistic case)
in applying the formalism of the relativistic `delta'-approximation
to the scattering in a point Coulomb field appears
due to the absence of the closed analytic expression for the scattering states
wavefunctions.
To overcome this difficulty we make another approximation and
use the Furry-Sommerfeld-Maue wavefunctions,
which are accurate up to the order $Z\alpha$ and for which
the closed analytical representation is known
(see, e.g.~\cite{Land4,Akhiezer,Gorshkov1964b,ElwertHaug69}).
The use of the FSM wavefunctions does not lead to essential difference
from the results obtained within the framework of the relativistic
distorted partial wave approach
if the main contribution to the cross section comes from the
partial waves of high orbital momentum  $l\gg Z\alpha$
\cite{Land4,Gorshkov1964b}.

So far there have been no numerical investigations of 
the domain of applicability of the FSM approximation to the 2BrS process. 
However, such an analysis was carried out for the single-photon 
BrS process \cite{ShafferPratt1997} and for the Compton
scattering of photons from bound electrons~\cite{BergstromSuricPiskPratt1993}.
It was established that for high $Z$ elements ($Z>60$) the 
calculations based on the FSM approximation underestimate 
the exact partial wave results for
double differential cross section by $10$--$50\%$
while for smaller $Z$ the difference is rarely greater than $20\%$
and typically is essentially less.
Theoretical estimations of the range of validity of the FSM approximation 
with respect to the atomic number $Z$ in connection 
with construction of the relativistic Green's
function can be found in~\cite{Gorshkov1964b,Hostler1964}.

The FSM wavefunction is given by (e.g.,~\cite{Land4}):
\begin{eqnarray}
\fl
\quad
\Psi^{(\sigma)}_{\nu}(\bfr)
=
  \ee^{\pi \xi/2}\,
\Gamma(1-\i\sigma\xi)\,
\ee^{\i\bfp\bfr/\hbar}
\left(
1 - {\i \hbar c \over 2 \E}\,\balpha \bnabla
\right)
F\Bigl( \i\sigma\xi, 1, -\i(\bfp\bfr - \sigma\, p r)/\hbar \Bigr)\,
u_{\nu}
\, .
\label{wf_FSM}
\end{eqnarray}
Here $\xi={Z\alpha\E}/p c$, the notation $F(a,b,z)$ stands for the
confluent hypergeometric function, $\Gamma(z)$ is the Gamma-function,
$u_{\nu}$ is the unit bispinor amplitude of a  plane wave.
In (\ref{wf_FSM}) and in what follows
the notations $\sigma$ is used for
(a) `$\pm$' if $\sigma$ is a superscript, and
(b) '$\pm 1$' if it is a factor.

Using (\ref{wf_FSM}) in (\ref{all2.2}) one derives the following
expression for the one-photon matrix element:
\begin{eqnarray}
\fl
\left[  \bfe\,\bfD^{(\Sb\Sa)}_{\nub \nua}(\bfk)  \right]_{\rm FSM}
&=&
C^{(\Sb\Sa)}_{p_{\rb} p_{\ra}}\, u_{\nub}^{\hc}\,
\Biggl[
\bfe\balpha\, I^{(\Sb\Sa)}_{\Pb\Pa}(\bfk)
\nonumber \\
\fl
&&
-
(\bfe\balpha)\, \Bigl( \balpha\bfI^{(\Sb\Sa)}_{\Pb\Pa}(\bfk) \Bigr)
-
\Bigl( \balpha[ \bfI^{(\Sa\Sb)}_{\,\Pa\Pb}(-\bfk)]^{*} \Bigr)
(\bfe\balpha)
 \Biggr]\, u_{\nua} \, .
\label{bD_FSM}
\end{eqnarray}
The factor $C^{(\Sb\Sa)}_{p_{\rb} p_{\ra}}$ and the integrals
$I^{(\Sb\Sa)}_{\Pb\Pa}(\bfk)$, $\bfI^{(\Sb\Sa)}_{\Pb\Pa}(\bfk)$
are defined as follows
\begin{eqnarray}
C^{(\Sb\Sa)}_{p_{\rb} p_{\ra}}
&=&
\ee^{\pi(\xib+\xia)/2}\,
\Gamma(1+\i\Sb \xib)\, \Gamma(1-\i\Sa \xia) \ ,
\label{def_C}\\
I^{(\Sb\Sa)}_{\Pb\Pa}(\bfk)
&=&
\int \d\bfr\,
\ee^{-\i(\Pb-\Pa+\hbar\bfk)\bfr/\hbar}\,
F^{(\Sb)\,*}_{\Pb}\, F^{(\Sa)}_{\Pa}\, ,
\label{def_I}\\
\bfI^{(\Sb\Sa)}_{\Pb\Pa}(\bfk)
&=&
{\i\hbar c \over 2 \E_{\ra}}\,
\int \d\bfr\,
\ee^{-\i(\Pb-\Pa+\hbar\bfk)\bfr/\hbar}\,
F^{(\Sb)\, *}_{\Pb}\,
\Bigl(\bnabla  F^{(\Sa)}_{\Pa}\Bigr)
\, .
\label{def_bI}
\end{eqnarray}
Here the short-hand notation
$F_{\bfp_{\rb,\ra}}^{(\sigma_{\rb,\ra})}\equiv
F\Bigl( \i \sigma_{\rb,\ra}\, \xi_{\rb,\ra} , 1,
-\i(\bfp_{\rb,\ra}\bfr - \sigma_{\rb,\ra} p_{\rb,\ra} r)/\hbar \Bigr)$
is used for the hypergeometric function.
The three integrals from (\ref{bD_FSM}) are related through~\cite{Land4}:
\begin{eqnarray}
q^2\, I^{(\Sb\Sa)}_{\Pb\Pa}(\bfk) =
{2 \E_{\rb} \over c}\,
\Bigl(\bfq\,\bfI^{(\Sa\Sb)}_{\Pa\Pb}(-\bfk)\Bigr)^{*}
-
{2 \E_{\ra} \over c}\,
\Bigl(\bfq\,\bfI^{(\Sb\Sa)}_{\Pb\Pa}(\bfk)\Bigr)
\label{int_rel}
\end{eqnarray}
where  $\bfq=\Pb-\Pa+\hbar\bfk$.
Following the procedure described in~\cite{Land4},
one transforms the term $\bfI^{(\Sb\Sa)}_{\Pb\Pa}(\bfk)$ to the
Nordsieck-type integral~\cite{Nordsieck} and then derives:
\begin{eqnarray}
\fl
\bfI^{(\Sb\Sa)}_{\Pb\Pa}(\bfk)
&=&
-{4 \pi\hbar^3 (Z \alpha) \over q^2}\,
{(q^2 - 2\Pb\bfq+ \i\Sb 0)^{\i\Sb\xib}
\over
(q^2 +2\Pa\bfq -\i\Sa 0)^{\i\Sa\xia+1}}
\, q^{2\i(\Sa\xia-\Sb\xib)}
\nonumber \\
\fl
&\times&
\Biggl\{
\bfq\,
F\Bigl(-\i\Sb \xib, \i\Sa\xia, 1, z\Bigr)
+
\i\Sb \xib\,
F\Bigl( 1 - \i\Sb \xib,  1 + \i\Sa\xia, 2,  z \Bigr)
\nonumber \\
\fl
&&
\times
\biggl[
{q^2 (\Pb-\Sa\Sb p_{\rb}\,\bfn_{\Pa})- 2\bfq (\Pb\bfq)
\over q^2-2\Pb\bfq}
- \bfq  z
\biggr]
\Biggr\} \, ,
\label{res_bI_a}
\end{eqnarray}
where
\begin{eqnarray}
z
=
2\,
{q^2 (\Pb\Pa - \Sa\Sb p_{\ra} p_{\rb}) - 2(\Pb\bfq)(\Pa\bfq)
\over
(q^2-2\Pb\bfq) (q^2+2\Pa\bfq)}\, .
\label{def_Z_for_Nord_int}
\end{eqnarray}

By analyzing the asymptotic form of the FSM wavefunction (\ref{wf_FSM})
one derives the following expression for the function
$f^{(\sigma)}_{\nub\nua}(\bfn,\bfe)$
(see~(\ref{A13c})):
\begin{eqnarray}
\left[ f^{(\sigma)}_{\nub\nua}(\bfn,\bfe) \right]_{\rm FSM}
=
\tilde{f}^{(\sigma)}_{\nub\nua}(\bfn,\bfe)
-
\sigma\,
{2 \pi\hbar \over \i p_{\ra}}
\, \delta(\bfn_{\Pa\!}-\sigma\bfn)\,
u_{\nub}^{\hc}\,  (\bfe\balpha)\, u_{\nua}
\, ,
\label{f_FSM}
\end{eqnarray}
where $\tilde{f}^{(\sigma)}_{\nub\nua}(\bfn,\bfe)$ is given by:
\begin{eqnarray}
\fl
\tilde{f}^{(\sigma)}_{\nub\nua}(\bfn,\bfe)
&=&
\hbar\,{\xia \over p_{\ra}}
{\Gamma(1-\i\sigma\xia) \over \Gamma(1+\i\sigma\xia)}
\,
{\ee^{\i\sigma\xia {\rm ln}((1-\sigma\cos{\theta_{\ra}})/2)}
\over 1 - \sigma\cos{\theta_{\ra}}}
\,
u_{\nub}^{\hc} (\bfe\balpha)\!
\left[
1 - {p_{\ra} c \over 2\E_{\ra}}\, \balpha(\bfn_{\Pa}-\sigma \bfn)
 \right]\! u_{\nua} \, ,
\label{tf_FSM}
\end{eqnarray}
with ${\cos{\theta_{\ra}} = \bfn_{\Pa}\bfn}$.

Now we are ready to write the amplitude $\cM^{(\delta)}$
(see (\ref{delta.12})) within the framework  of the FSM approximation.
For the negative-energy part one obtains:
\begin{eqnarray}
\fl
\left[{\cal M}_{[12]}\right]_{-}
\!
+
\left[{\cal M}_{[21]}\right]_{-}
\!=\!
-
{\bfe_1 \bfe_2 \over m_{\rm p} c^2}\, C^{(-+)}_{p_2 p_1}\,
u_{\nu_2}^{\hc}\!
\left[
I^{(-+)}_{\bfp_2\bfp_1}(\bfk)
-
\balpha\bfI^{(-+)}_{\bfp_2\bfp_1}(\bfk)
-
\balpha\bfI^{(+-)\,*}_{\bfp_1\bfp_2}(-\bfk)\right] u_{\nu_1}
\label{SG_FSM}
\end{eqnarray}
where $\bfk=\bfk_1+\bfk_2$.

The FSM expression for the sum $I = I_{[12]}+I_{[21]}$ reads
\begin{eqnarray}
\fl
I
=
e_{1 i} e_{2 j}
\, u_{\nu_2}^{\hc}
\Biggl\{
{p C^{(-\sigma)}_{p_2 p}  C^{(\sigma +)}_{p  p_1}
\over 16\pi^2 \hbar^3 c^2}\!\!
\int
\d \Omega_{\bfp}\!
\left[
\alpha_j  I^{(-\sigma)}_{\,\bfp_2\bfp}(\bfk_2)
-
\alpha_j\,\balpha\bfI^{(-\sigma)}_{\bfp_2\bfp}(\bfk_2)
-
\balpha\bfI^{(\sigma -)\,*}_{\bfp\bfp_2}(-\bfk_2)\, \alpha_j
 \right]
\nonumber\\
\fl
\quad
\times
(c \balpha \bfp + \E_p + \gamma^0 m_{\rm p} c^2)
\biggl[
\alpha_i  I^{(\sigma +)}_{\bfp\bfp_1}(\bfk_1)
-
\alpha_i \balpha\bfI^{(\sigma +)}_{\bfp\bfp_1}(\bfk_1)
-
\balpha\bfI^{(+\sigma)*}_{\bfp_1\bfp}(-\bfk_1) \alpha_i
 \biggr]
\Biggr]_{\E_p=\E_1-\hbar\om_1}\!\!\!\!\! u_{\nu_1}
\label{I12_FSM}
\end{eqnarray}
where the subscripts $i, j$ denote the Cartesian coordinates,
and the rule which is adopted is $a_ib_i = {\bf ab}$.

Finally, using (\ref{bD_FSM}) and (\ref{f_FSM}) in
(\ref{delta.8}), and carrying out the summation over $\mu$ one writes:
\begin{eqnarray}
\fl
R_{[12]}^{(\delta)}
&=&
{1 \over 4}
\, e_{1 i} e_{2 j}
\, u_{\nu_2}^{\hc}
\Biggl\{
\Bigl[
{\cal R}_{[12]}^{({\rm I})}(\bfp_1 - \hbar\bfk_1) \Bigr]_{ij}
+
\Bigl[{\cal R}_{[12]}^{({\rm II})}(\bfp_2 + \hbar\bfk_2) \Bigr]_{ij}
\Biggr.
\nonumber \\
\fl
&&
+
{\i  \xi_1 \over 2 \pi}
{\Gamma(1 - \i\xi_1) \over \Gamma(1 +\i\xi_1)}
\int_0^{\pi}  \sin\theta_{\bfp}\, \d \theta_{\bfp}
\int_0^{2\pi} \d\phi_{\bfp}\,
\Bigl[  {\cal R}_{[12]}^{({\rm III})}(p_1\bfn_{\bfp}-\hbar\bfk_1) \Bigr]_{ij}
\nonumber \\
\fl
&&
+
{\i  \xi_2 \over 2 \pi}
{\Gamma(1 - \i\xi_2) \over \Gamma(1 + \i\xi_2)}
\int_0^{\pi}  \sin\theta_{\bfp}\, \d\theta_{\bfp}
\int_0^{2\pi}\d\phi_{\bfp}\,
\Bigl[  {\cal R}_{[12]}^{({\rm IV})}(p_2\bfn_{\bfp}+ \hbar\bfk_2) \Bigr]_{ij}
\Biggr\}
u_{\nu_1}\, .
\label{R12_FSM}
\end{eqnarray}
Here $\theta_{\bfp},\phi_{\bfp}$ are the polar angles
of the vector $\bfp$ of the intermediate state.
The tensors $\Bigl[{\cal R}_{[12]}^{(\dots)}\Bigr]_{ij}$
are defined as follows
\begin{eqnarray}
\fl
\Bigl[  {\cal R}_{[12]}^{({\rm I})}(\bfs) \Bigr]_{ij}
&=&
\Biggl\{
\alpha_j
\biggl[
C^{(-+)}_{p_2 s} I^{(-+)}_{\bfp_2 \bfs}(\bfk_2) +
B^{(--)}_{p_2 s}(\xi_1)I^{(--)}_{\bfp_2 \bfs}(\bfk_2)
 \biggr]
\nonumber \\
\fl
&&
-
\alpha_j \balpha
\biggl[
C^{(-+)}_{p_2 s} \bfI^{(-+)}_{\bfp_2 \bfs}(\bfk_2)
+
B^{(--)}_{p_2 s}(\xi_1) \bfI^{(--)}_{\bfp_2 \bfs}(\bfk_2)
 \biggr]
\nonumber \\
\fl
&&
-
\biggl[
C^{(-+)}_{p_2 s} \bfI^{(+-)*}_{\bfs\bfp_2}(-\bfk_2)
+
B^{(--)}_{p_2 s}(\xi_1) \bfI^{(--)*}_{\bfs\bfp_2}(-\bfk_2)
 \biggr]
\balpha \alpha_j
 \Biggr\}
  A(\bfs,\om_1)  \alpha_i
\,,
\label{R(I)}
\end{eqnarray}
%%%%%%%%%%%%%%%
\begin{eqnarray}
\fl
\Bigl[  {\cal R}_{[12]}^{({\rm II})}(\bfs) \Bigr]_{ij}
&=&
\alpha_j A(\bfs,\om_1)
\Biggl\{
\alpha_i
\biggl[
C^{(-+)}_{s p_1}I^{(-+)}_{\bfs\bfp_1}(\bfk_1)
+
B^{(++)}_{s p_1}(\xi_2) I^{(++)}_{\bfs\bfp_1}(\bfk_1)
 \biggr]
\nonumber \\
\fl
&&
-
\alpha_i \balpha
\biggl[
C^{(-+)}_{s p_1}\bfI^{(-+)}_{\bfs\bfp_1}(\bfk_1)
+
B^{(++)}_{s p_1}(\xi_2) \bfI^{(++)}_{\bfs\bfp_1}(\bfk_1)
\biggr]
\nonumber \\
\fl
&&
-\biggl[
C^{(-+)}_{s p_1}\bfI^{(+-)*}_{\bfp_1 \bfs}(-\bfk_1)
+
B^{(++)}_{s p_1}(\xi_2)\bfI^{(++)*}_{\bfp_1 \bfs}(-\bfk_1)
\biggr]
\balpha \alpha_i
  \Biggr\}
\,,
\label{R(II)}
\end{eqnarray}
%%%%%%%%%%%%%
\begin{eqnarray}
\fl
\Bigl[  {\cal R}_{[12]}^{({\rm III})}(\bfs) \Bigr]_{ij}
\!
&=&
C^{(--)}_{p_2 s}
{ \ee^{\i\xi_1 \ln[(1-\cos{\theta_1})/2]} \over  1-\cos{\theta_1} }
\biggl[
\alpha_j  I^{(--)}_{\bfp_2 \bfs}(\bfk_2)
 -
\alpha_j   \balpha\bfI^{(--)}_{\bfp_2 \bfs}(\bfk_2)
-
\balpha\bfI^{(--)*}_{\bfs\bfp_2}(-\bfk_2) ]\alpha_j
 \biggr]
\nonumber \\
\fl
&&
\times
A(\bfs,\om_1)
\alpha_i
\left[1 - {c \over 2 \E_1}\, \balpha (\bfp_1 - \bfs - \hbar\bfk_1) \right]
\,,
\label{R(III)}
\end{eqnarray}
%%%%%%%%%%%%%
\begin{eqnarray}
\fl
\Bigl[  {\cal R}_{[12]}^{({\rm IV})}(\bfs) \Bigr]_{ij}
&=&
C^{(++)}_{s p_1}
{ \ee^{\i\xi_2 \ln[(1-\cos{\theta_2})/2]} \over 1-\cos{\theta_2} }
\left[
1 - {c \over 2 \E_2}\,\balpha (\bfp_2 - \bfs + \hbar\bfk_2)
 \right]
\alpha_j
A(\bfs,\om_1)
\nonumber \\
\fl
&&
\times
\left[
\alpha_i  I^{(++)}_{\bfs\bfp_1}(\bfk_1)
-
\alpha_i  \balpha\bfI^{(++)}_{\bfs\bfp_1}(\bfk_1)
-
\balpha\bfI^{(++)*}_{\bfp_1\bfs}(-\bfk_1)  \alpha_i
 \right]
\,.
\label{R(IV)}
\end{eqnarray}
Here
 $\cos{\theta_1} =\bfp_1 (\bfs+\hbar\bfk_1)/p_1^2$
and
$\cos{\theta_2} =\bfp_2 (\bfs-\hbar\bfk_2)/p_2^2$
is used in (\ref{R(III)}) and (\ref{R(IV)}).
The quantities $B^{(\Sb\Sa)}_{p_{\rb} p_{\ra}}(\xi)$ and $A(\bfp,\om)$
are given by
\begin{eqnarray*}
B^{(\Sb\Sa)}_{p_{\rb} p_{\ra}}(\xi)
=
{\Gamma(1 - \i \xi) \over \Gamma(1 + \i \xi)}
\, C^{(\Sb\Sa)}_{p_{\rb} p_{\ra}} \, ,
\quad
A(\bfp,\om)
=
{c \balpha \bfp + \E_p + \gamma^0 m_{\rm p} c^2
\over \E_p(\E_p + \hbar\om - \E_1)}
\label{def_A}
\end{eqnarray*}
with
$\E_p= \sqrt{p^2 c^2 + m_{\rm p}^2 c^4} $.

%%%%%%%%%%%%%%%%%%%%%%%%%%%%%%%%%%%%%%%%%%%%%%%%%%%%%%%%%%%%%%%%%%%%%%%%%
\section{Numerical results}
\label{results}
%%%%%%%%%%%%%%%%%%%%%%%%%%%%%%%%%%%%%%%%%%%%%%%%%%%%%%%%%%%%%%%%%%%%%%%%%

\begin{figure}
\begin{center}
\includegraphics[width=12cm,height=13cm,angle=270]{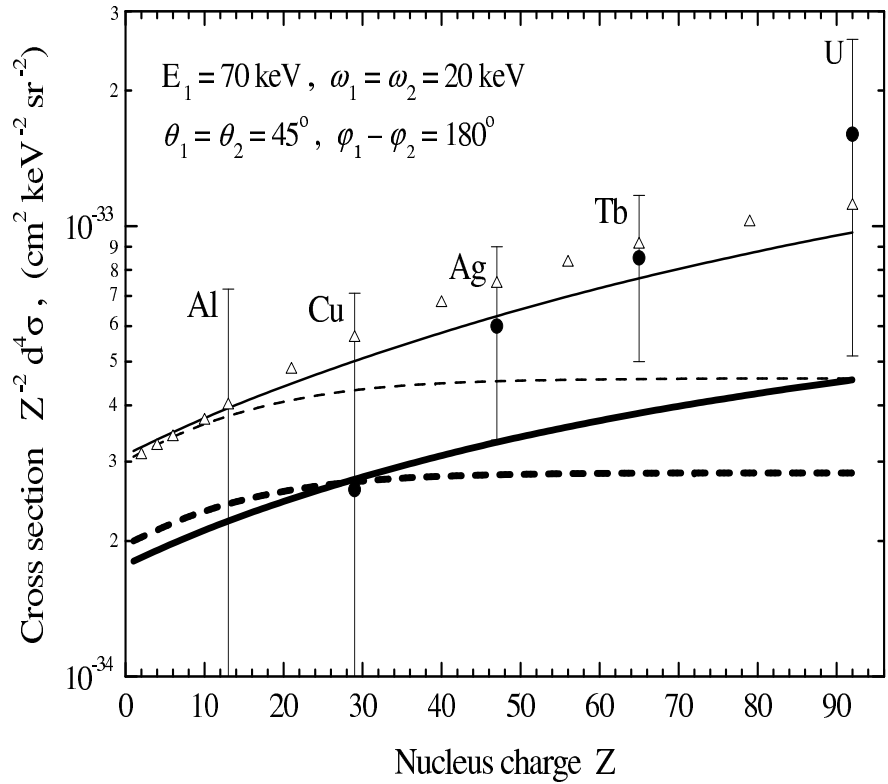}
\end{center}
\caption{
Dependence of $Z^{-2}\, \d^4\sigma_{\om_1\om_2}(\Om_1,\Om_2)$ on the nucleus
charge $Z$.
The incoming electron energy, the energies of the photons
and the geometry of the emission are as indicated.
Filled circles with the error bars are the experimental data
from~\cite{Kahler92}.
Solid lines correspond to the relativistic (thick line)
and non-relativistic (thin line) `delta'-approximations.
Dashed lines are used for the dependences obtained within
the framework of relativistic (thick line)
and non-relativistic (thin line) Born-Elwert approximations.
Triangles present the results of the exact non-relativistic non-dipole
approximation~\cite{Korol1997}.}
\label{fig_Z-dep_1}
\end{figure}

\begin{figure}
\begin{center}
\includegraphics[width=12cm,height=13cm,angle=270]{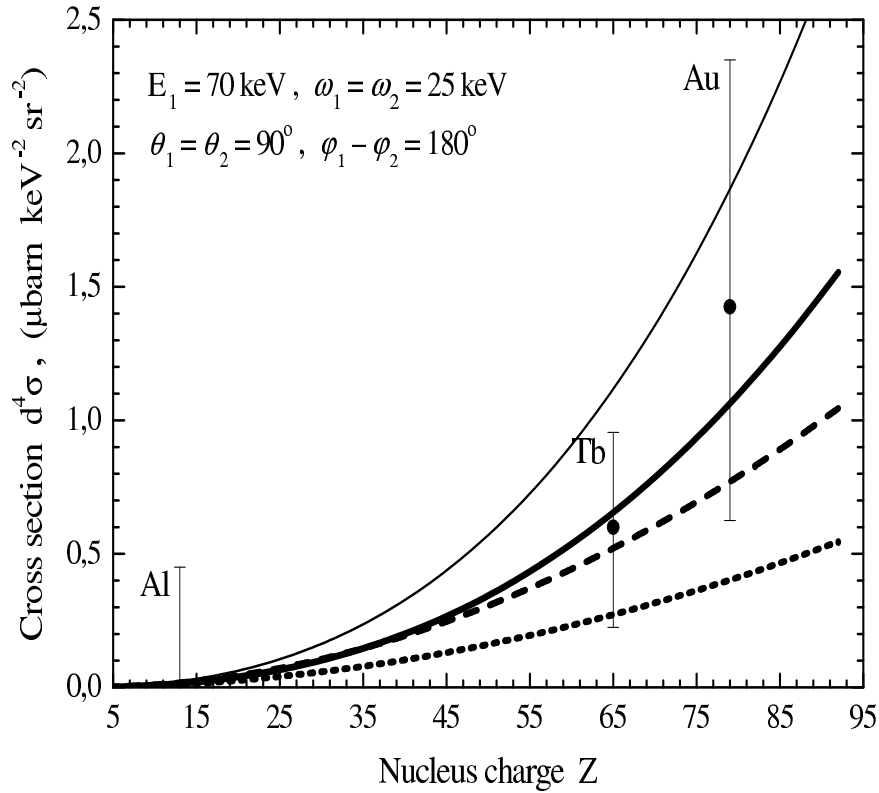}
\end{center}
\caption{
Calculated cross section $\d^4\sigma_{\om_1\om_2}(\Om_1,\Om_2)$ as a
function of $Z$ versus the experimental data from~\cite{Quarles93a}.
Solid lines correspond to the relativistic (thick lines)
and non-relativistic (thin lines) `delta'-approximations.
The results of relativistic plane-wave Born  and
relativistic Born-Elwert approximations are drawn with
the short-dashed and the long-dashed lines, respectively.
The energies $\E_1$, $\om_{1,2}$ and the emission geometry are as
indicated.}
\label{fig_Z-dep_2}
\end{figure}

\begin{figure}
\begin{center}
\includegraphics[width=12cm,height=13cm,angle=270]{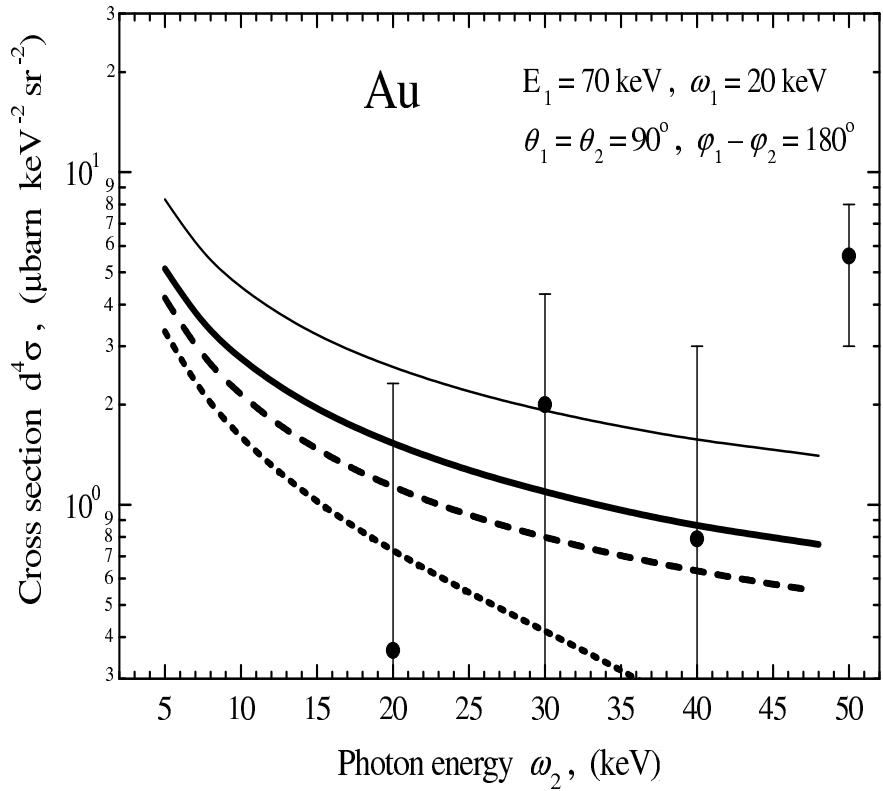}
\end{center}
\caption{
The calculated cross section
$\d^4\sigma_{\om_1\om_2}(\Om_1,\Om_2)$
as a function of $\om_2$ versus the experimental
data~\cite{Quarles93a} for a 70 keV  electron collision with an
Au atom ($Z=79$).
Solid lines correspond to the relativistic (thick line)
and non-relativistic (thin line) `delta'-approximations.
Short-dashed line represents the relativistic plane-wave Born
approximation,
the long-dashed one stands for the correction due to the Elwert factor
(\protect\ref{ElwertBorn}).}
\label{fig_w2-dep_1}
\end{figure}

\begin{figure}
\begin{center}
\includegraphics[width=12cm,height=13cm,angle=270]{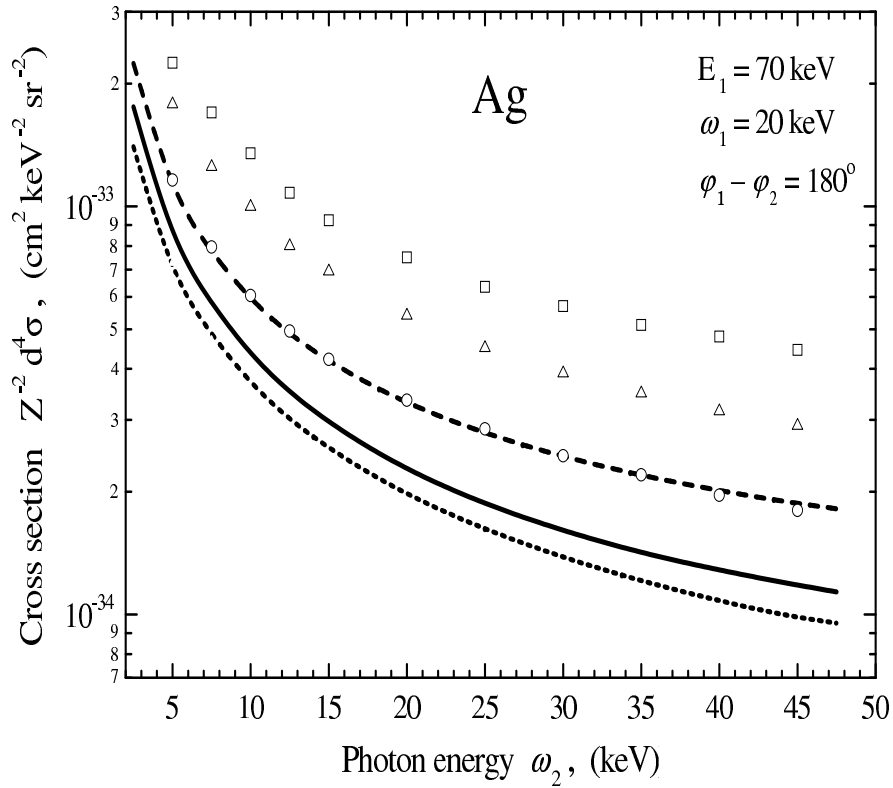}
\end{center}
\caption{
Results from the relativistic `delta'-approximation theory versus
the exact non-relativistic calculations~\cite{Korol1997}
for $Z^{-2}\, \d^4\sigma_{\om_1\om_2}(\Om_1,\Om_2)$
in a 70 keV electron collision with an Ag atom (${Z=47}$).
The sets of curves (the `delta'-approximation) and symbols
(the exact non-relativistic results) correspond to different
emission geometries:
(a) solid line and triangles are used for
$\theta_1=\theta_2=30^{\circ}$,
(b) long-dashed line and squares for $\theta_1=\theta_2=45^{\circ}$,
and
(c) short-dashed line and circles for $\theta_1=\theta_2=90^{\circ}$.}
\label{fig_w2-dep_2}
\end{figure}

\begin{figure}
\begin{center}
\includegraphics[width=12cm,height=13cm,angle=270]{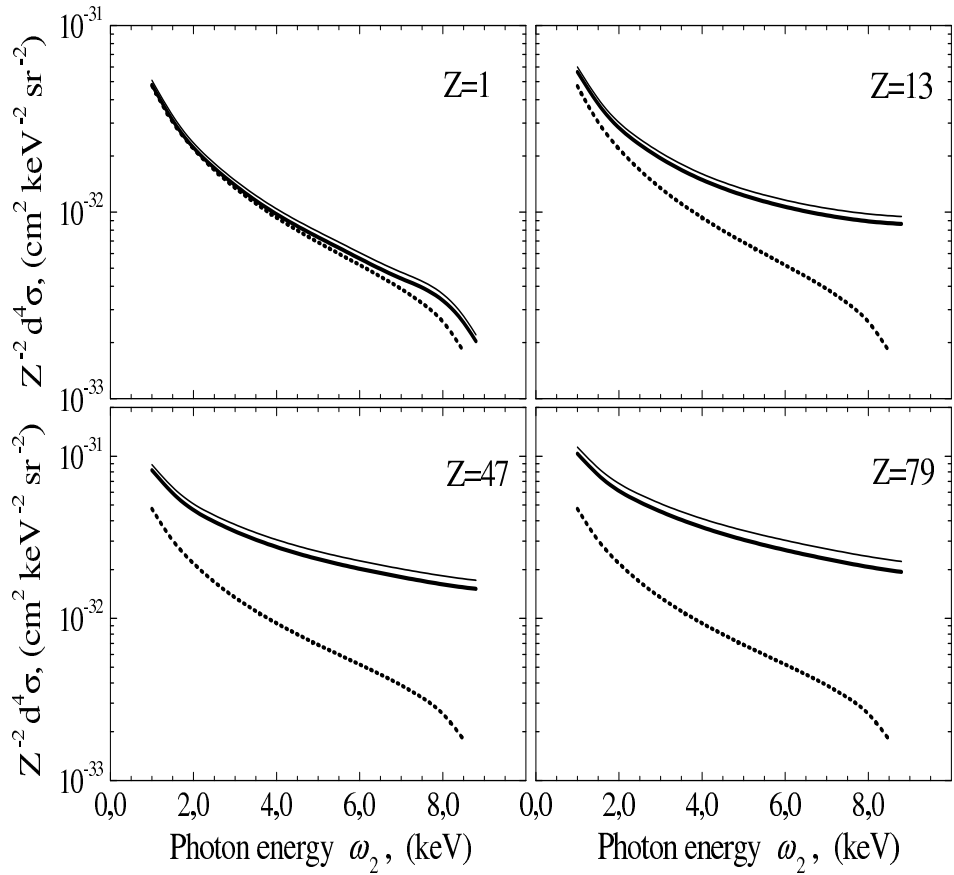}
\end{center}
\caption{
Comparison of the results of relativistic `delta'-approximation with
the exact non-relativistic calculations with accounting for radiation
retardation~\cite{DonderaFlorescu1998} of 2BrS cross section
$Z^{-2}\, \d^4\sigma_{\om_1\:\!\om_2}(\Om_1,\Om_2)$
as function of energy of second emitted photon $\om_2$.
Presented four graphs correspond to collision of $10$~keV electron
with four different atomic targets with ${Z=1,\, 13,\, 47,\, 79}$.
The energy of the first emitted photon $\om_1=1$~keV and emission
geometry $\theta_1=\theta_2=45^{\rm o}$, $\varphi_1-\varphi_2=180^{\rm o}$
are the same for all graphs.
Results of relativistic `delta'-approximation are given by thick solid line.
For results of non-relativistic exact calculations the thin solid line is
used. Relativistic plane-wave Born curves are plotted by dashed line.}
\label{fig_Z-dep_Pratt}
\end{figure}

The approach described in sections~\ref{delta} and~\ref{FSM}
was applied to calculate the spectral-angular distribution
of the 2BrS formed in
an electron scattering in Coulomb fields of a variety of charges.
The figures correspond to the incoming electron kinetic energy
$\E_1=70$ keV, except for figure~\ref{fig_Z-dep_Pratt} where
$\E_1=10$ keV.
In figures~\ref{fig_Z-dep_1},~\ref{fig_w2-dep_1} and~\ref{fig_w2-dep_2}
the energy of the first photon is fixed at $\hbar\omega_1=20$ keV,
while for the second photon it varies within the range
$\hbar\om_2=2.5\dots 47.5$ keV.
In figures~\ref{fig_Z-dep_2} and~\ref{fig_Z-dep_Pratt}
the curves are plotted for $\hbar\om_1=\hbar\om_2=25$ keV and
$\hbar\om_1=1$ keV,  $\hbar\om_2=1\dots 8.5$ keV, respectively.
%%%%%%%%%%%%%%
%%%%%%%%%%%%%%%%%%%%%%%%%%%%%%%%%%%%%%%%%%%%%%%%%%%%%%%%%%%%%%%%%%%%%%%%%%%

The results in all graphs refer to the com planar
geometry when the vectors $\bfp_1$, $\bfk_1$ and $\bfk_2$
lay in the same plane.
The emission angles  $(\theta_{1,2},\varphi_{1,2})$,
are measured with respect $\bfp_1$.
Because of the axial symmetry the differential cross section
$\d^4\sigma/(\d\om_1 \d\om_2 \d\Om_1 \d\Om_2)
\equiv \d^{4}\sigma_{\om_1 \om_2}(\Om_1,\Om_2)$ depends
on the difference $(\varphi_1-\varphi_2)$
but not on the angles $\varphi_1$ and $\varphi_2$ separately.

The kinetic energy  $\E_1=70$ keV, the photon energies and the geometry
of the radiation correspond to the experimental conditions
\cite{Kahler92,Quarles93a,Quarles93b,Quarles93c}.

%%%%%%%%%%%%%%%%%%%%%%%%%%%%%%%%%%%%%%%%%%%%%%%%%%%%%%%%%%%%%%%%%%%%%%%%%%%
%%%%%%%%%%%%%%%%%%%%%%%%%%%%%%%%%%%%%%%%%%%%%%%%%%%%%%%%%%%%%%%%%%%%%%%%%%%

%%%%%%%%%%%%%%%%%%%%%%%%%%%%%%%%%%%%%%%%%%%%%%%%%%%%%%%%%%%%%%%%%%%%%%%%%%%
%%%%%%%%%%%%%%%%%%%%%%%%%%%%%%%%%%%%%%%%%%%%%%%%%%%%%%%%%%%%%%%%%%%%%%%%%%%

Numerical calculation of the relativistic amplitudes was
carried out within the framework of the `delta'-approximation and
using the FSM wavefunctions.
The analogous non-relativistic calculations were performed
with the exact Coulomb wavefunctions.
The results of relativistic Born calculations presented in the
figures were obtained by programming the formulae
presented in~\cite{FedorovaKorolSolovjev2000}.
The curves corresponding to the relativistic Born-Elwert
approximation represent the relativistic Born curves
corrected by the Elwert factor~\cite{ElwertHaug69}
(we used its relativistic analogue proposed in~\cite{AvdPratt99}):
\begin{eqnarray}
f_{\rm Elw} =
{p_1 \over p_2}\,
{1 - \exp{(-2 \pi \xi_1)} \over
1 - \exp{(-2 \pi \xi_2)}}
\label{ElwertBorn}
\end{eqnarray}

%%%%%%%%%%%%%%%%%%%%%%%%%%%%%%%%%%%%%%%%%%%%%%%%%%%%%%%%%%%%%%%%%%%%%%%%%%%
%%%%%%%%%%%%%%%%%%%%%%%%%%%%%%%%%%%%%%%%%%%%%%%%%%%%%%%%%%%%%%%%%%%%%%%%%%%

In figures~\ref{fig_Z-dep_1}--\ref{fig_w2-dep_1} we compare the results
of calculation of 2BrS cross section
obtained within the framework of relativistic `delta'-approximation with
the available experimental data
\cite{Kahler92,Quarles93a,Quarles93b,Quarles93c} and the
the results from other theories.
The latter include the exact non-relativistic non-dipole treatment of the
2BrS process in a point Coulomb field~\cite{Korol1997,DonderaFlorescu1998},
the non-relativistic non-dipole `delta'-approximation~\cite{Korol1996}, and
the relativistic plane-wave Born approximation with and without
the correction factor (\ref{ElwertBorn}).
The calculations within the framework of the non-relativistic dipole approximations,
although having been carried out, are not presented here.
This is because, as it is known~\cite{Korol1996,DonderaFlorescuPratt1996},
the dipole-photon scheme, which neglects the correction terms of the
leading order $v/c$, strongly underestimates the magnitude of the 2BrS cross section.

One of the conclusions which can be drawn on the basis of
the data presented in figures~\ref{fig_Z-dep_1}--\ref{fig_w2-dep_1} is that
the relativistic effects result in a decrease  of the magnitude of the
cross section.
This is clearly seen if one compares the dependences
obtained by using the relativistic (thick line) and non-relativistic
(thin line) `delta'-approximations, which are presented in  figure~\ref{fig_Z-dep_1}.
Qualitatively, the influence of the relativistic corrections could be illustrated
in terms of classical electrodynamics as follows.
Apart from the spin-related effects, the general consequence of a movement
with relativistic velocity is the dependence of mass of a projectile on its
velocity, which leads to the increase of the mass as $m=m_0\,(1  - v^2/c^2)^{-1/2}$
($m_0$ is the rest mass).
This effect reduces the magnitude of the projectile's acceleration, which
defines the intensity of radiation.

%%%%%%%%%%%%%%%%%%%%%%%%%%%%%%%%%%%%%%%%%%%%%%%%%%%%%%%%%%%%%%%%%%%%%%%%%%%
%%%%%%%%%%%%%%%%%%%%%%%%%%%%%%%%%%%%%%%%%%%%%%%%%%%%%%%%%%%%%%%%%%%%%%%%%%%

Presented results show that the relativistic `delta'-approximation, on the whole,
reproduces quite well the available experimental data.
However, there are two exceptions.
In figure~\ref{fig_Z-dep_1} in the range $Z\sim 100$
the calculated values underestimate the experimental data.
The possible explanation of this discrepancy is in the inadequacy of the
FSM approach in the limit $\alpha Z\sim 1$,
The second example of a strong deviation between the theory and the experiment
is seen in figure~\ref{fig_w2-dep_1} at the tip-end of the spectrum.
However, in this case, the discrepancy may be due to the experimental error,
since the experimental data for $\om_2=50$ keV lies far above all theoretical
predictions.

In figure~\ref{fig_w2-dep_2} we compare the predictions obtained within
relativistic `delta'-approximation scheme with the results of the
exact non-relativistic theory  (the retardation included)~\cite{Korol1997}.
It was demonstrated in the cited paper in the non-relativistic domain
the `delta'-approximation is in a good agreement with the exact calculations.
The account for the relativistic effects leads to the decrease
in the cross section magnitude.
This influence has been already mentioned above.
We note also that this decrease is dependent on the geometry of the emission.
The nature of this feature is not fully clear at present,
and we believe to get the explanation as soon as the
exact relativistic results will become available.

For the lower values of the  projectile energy
the influence of the relativistic corrections is less pronounced.
This is demonstrated by figure~\ref{fig_Z-dep_Pratt} where the results of
the relativistic `delta'-approximation (thick solid lines) for a 10 keV electron
are compared with the exact non-relativistic non-dipolar calculations
(thin solid lines) presented in~\cite{DonderaFlorescu1998}.
Let us point out that the non-relativistic theory based on the `delta'-approximation
produces the results (not plotted in the figure)
which practically coincide with the thin lines for all elements presented.
The difference between the thick and thin solid curves increases with the atomic number
$Z$.
Although the exact nature of the difference can be established when the
exact relativistic theory of the effect becomes available, some part of it can
be attributed to the use of the FSM wavefunctions which accuracy also decreases
with $Z$.

For the sake of comparison in figure~\ref{fig_Z-dep_Pratt} we present the
curves corresponding to the relativistic plane-wave  Born approximation
(dotted lines).
It is clearly seen that the simpler theory becomes absolutely inapplicable
in the range of medium and large $Z$ values.

On the whole, we may state that the developed approach is adequate
for the description of 2BrS process and is an efficient tool for numerical
analysis of the characteristics of the process
(the spectral and spectral-angular intensities  of the radiation).
The results obtained within the `delta'-approximation
(both relativistic and non-relativistic) describe quite well the
behaviour of 2BrS cross sections over wide ranges of energies
of the projectile electron and the photons and of atomic numbers $Z$.
The influence of the relativistic effects increases  with $\E_1$ and $Z$, and
they must be accounted for in addition to the radiation retardation effect.

%%%%%%%%%%%%%%%%%%%%%%%%%%%%%%%%%%%%%%%%%%%%%%%%%%%%%%%%%%%%%%%%%%%%%%%%%%%
\section{Conclusions}
\label{conclusions}
%%%%%%%%%%%%%%%%%%%%%%%%%%%%%%%%%%%%%%%%%%%%%%%%%%%%%%%%%%%%%%%%%%%%%%%%%%%

Basing on the numerical results presented above, we regard the developed
approach as more reliable than the simpler relativistic theories.
Within the framework of the `delta'-approximation  the analytical
structure of the 2BrS amplitude is simplified considerably.
Additionally, this method allowed, for the first time, to carry out
numerical analysis  of the two-photon free-free transitions in relativistic domain
beyond the plane wave Born approximation.
The method is easily generalized to the case of a $n$-photon free-free transition.

Nevertheless, despite the (relative) simplicity and the efficiency of the
outlined formalism, by no means does it solve in full
the problem of the free-free relativistic transitions.
The exact relativistic treatment of the process is still unavailable.
This task is difficult from  the analytical and the computational viewpoints.
The direct evaluation of the amplitude by means of the partial wave
expansion of the initial/final states wavefunction, the Green's function
and the multipole expansions for the emitted photons
seems hardly to be implemented in the nearest future.

However, we are more optimistic on the prospects of two other approaches.
The first one is based on the exact relativistic treatment of the 2BrS
process within the framework of the FSM approximation, where one can construct,
in addition to the wavefunctions, a practically useful form for the Green's function
(see~\cite{Gorshkov1964b}).
The second approach is based on the Sternheimer-Dalgarno-Lewis method
\cite{Sternheimer,DalgarnoLewis} for the calculation of two-photon transition
amplitude between two states of a continuous spectrum.
The work in these directions is being carried out.

%%%%%%%%%%%%%%%%%%%%%%%%%%%%%%%%%%%%%%%%%%%%%%%%%%%%%%%%%%%%%%%%%%%%%%%%%%%%
\ack
%%%%%%%%%%%%%%%%%%%%%%%%%%%%%%%%%%%%%%%%%%%%%%%%%%%%%%%%%%%%%%%%%%%%%%%%%%%%

This work was supported in part by the grant of the Ministry of Education
of Russian Federation
and by the Russian Foundation for Basic research (grant No 03-02-16415-a).
AVK acknowledges the support from Alexander von Humboldt Foundation.
The authors are grateful to Viorica Florescu for sending the copy of
the paper~\cite{GhilenceaToaderDiaconu1995}.

%%%%%%%%%%%%%%%%%%%%%%%%%%%%%%%%%%%%%%%%%%%%%%%%%%%%%%%%%%%%%%%%%%%%%%%%%%%%
%%%%%%%%%%%%%%%%%%%%%%%%%%%%%%%%%%%%%%%%%%%%

%%%%%%%%%%%%%%%%%%%%%%%%%%%%%%%%%%%%%%%%

\section*{References}

\begin{thebibliography}{99}

%%%%%%%
\bibitem{Korol1994a}
        Korol, A.~V.,
        1994.
        J.~Phys.~B  27, 155-174.
%%%%%%%
\bibitem{Korol1995a}
        Korol, A.~V.,
        1995.
        J.~Phys.~B 28, 3873-3887.
%%%%%%%
\bibitem{Korol1996}
        Korol, A.~V.,
        1996.
        J.~Phys.~B 29, 3257-3276.
%%%%%%%
\bibitem{Korol1997}
        Korol, A.~V.,
        1997.
        J.~Phys.~B  30, 413-438.
%%%%%%%%%
\bibitem{FedorovaKorolSolovjev2000}
        Fedorova, T.~A., Korol, A.~V., Solovjev, I.~A.,
        2000.
        J.~Phys.~B 33, 5007-5024.
%%%%%%%
\bibitem{MaquetVeniardMarian1998}
        A.~Maquet. A., V\'eniard, V., Marian, T.~A.,
        1998.
        J.~Phys.~B 31, 3743-3764.
%%%%%%%%%
\bibitem{EhlotzkyJaronKaminski1998}
        Ehlotzky, F., Jaro\'n, A., Kamis\'nki, J.~Z.,
        1998.
        Phys. Rep. 297,  63-153.
%%%%%%%%
\bibitem{KorolSolovyov1997}
        Korol, A.~V., Solov'yov, A.~V.,
        1997.
        J.~Phys.~B 30, 1105-1150.
%%%%%%%%
\bibitem{Land4}
       Berestetskii, V.~B., Lifshitz, E.~M., Pitaevskii, L.P.,
       1982.
       Quantum Electrodynamics. Pergamon, Oxford.
%%%%%%%
\bibitem{Akhiezer}
        Akhiezer, A.~I., Berestetskii V.~B.,
        1969.
        Quantum Electrodynamics. Nauka, Moscow.
%%%%%%% ++
\bibitem{Rosenberg1994}
         Rosenberg, L.,
         1994. Phys.~Rev.~A 49, 4770-4777.
%%%%%%%
\bibitem{Korol1995b}
        Korol, A.V.,
        1995.
        Nucl. Instrum. Methods B 99, 160-162.
%%%%%%%
%\bibitem{GrIvKoKu99}
%        Gribakin~G.F., Ivanov~V.K., Korol~A.V., and Kuchiev~M.Yu.
%Three-photon detachment of electrons from the fluorine negative ion//
%        J.Phys.B: At.Mol.Opt.Phys. -- 1999 -- v.32 -- p.5463-5478

%%%%%%%
%\bibitem{GrIvKoKu00}
%        Gribakin~G.F., Ivanov~V.K., Korol~A.V., and Kuchiev~M.Yu.
%Two-photon detachment of electrons from halogen negative ions//
%        J.Phys.B: At.Mol.Opt.Phys. -- 2000 -- v.33 -- p.821-828

%%%%%%%% ++
\bibitem{FlorescuDjamo1986}
        Florescu, V., Djamo, V.,
        1986.
        Phys. Lett. 119A, 73-76.

%%%%%%%% ++
\bibitem{VeniardGavrilaMaquet1987}
        V\'eniard, V., Gavrila, M., Maquet, A.,
        1987.
        Phys.~Rev.~A 35, 448-451.
%%%%%%%% ++
\bibitem{GavrilaMaquetVeniard1990}
        Gavrila, M., Maquet, A., V\'eniard ~V.,
        1990.
        Phys.~Rev. ~A, 42, 236-247.
%%%%%%%%
\bibitem{DonderaFlorescu1993}
        Dondera, M., Florescu, V.,
        1993.
        Phys.~Rev.~A 48, 4267-4271.
%%%%%%%%
\bibitem{DonderaFlorescu1998}
        Dondera, M., Florescu, V.,
        1998.
        Phys.~Rev.~A 58, 2016-2022.
%%%%%%% ++
\bibitem{VeniardPiraux1990}
        V\'eniard, V.,  Piraux, B.,
        1990.
        Phys.~Rev.~A 41, 4019-4034.
%%%%%%%% ++
\bibitem{BlahaDavis1995}
        Blaha, M.,  Davis, J.
        1995.
        Phys.~Rev.~A 51, 2308-2315.
%%%%%%%% ++
\bibitem{MercourisKomninosDionissopoulouNicolaides1996}
        Mercouris, N., Komninos, Y., Dionissopoulou, S., Nicolaidest, C.~A.
        1996.
        J.~Phys.~B 29, L13-L19.
%%%%%%%
\bibitem{GribakinIvanovKorolKuchiev1999_2000}
        Gribakin G.~F., Ivanov V.~K., Korol~A.~V., Kuchiev~M.~Yu.,
        1999.
        J.~Phys.~B 32, 5463-5478;
        2000. ibid. 33, 821-828.
%%%%%%%%
\bibitem{Korol1994c}
        Korol, A.~V.,
        1994.
        J.~Phys.~B 27, 4765-4777.
%%%%%%% ++
\bibitem{ManakovMarmoFainstein1995}
        Manakov, N. L., Marmo, S. I., Fainstein, A. G.,
        1995.
        JETP 81, 860 -877.
%%%%%%% ++
\bibitem{KrylovetskiiManakovMarmoStarace2002}
        Krylovetskii, A. A., Manakov, N. L., Marmo, S. I., Starace, A. F.
        2002.
        JETP 95, 1006-1032.
%%%%%%%% ++
\bibitem{FlorescuFlorescu2000}
        Florescu, A., Florescu, V.,
        2000.
        Phys.~Rev. A 61, 033406.

%%%%%%%%%% Experiments
%%%%%%%%
\bibitem{AltmanQuarles1985}
        Altman, J. C.,  Quarles, C. A.,
        1985,
        Phys. Rev. A 31,  2744;
        1985, Nucl. Instrum. Methods A 240,  538. 

%%%%%%%
\bibitem{LehtihetQuarles1989}
        Lehtihet, H. E., Quarles, C.  A.,
        1989,
        Phys. Rev. A 39, 4274.

%%%%%%%
\bibitem{Hippler1991}
        Hippler, R.,
        1991,
        Phys. Rev. Lett. 66, 2197.

%%%%%%%
\bibitem{Kahler92}
         Kahler, D.~L., Liu, J., Quarles, C.~A.,
         1992.
         Phys. Rev. A  45,  R7663-R7666;
         1992. Phys. Rev. Lett. 68, 1690-1693.
\bibitem{Quarles93a}
        Liu, J., Quarles, C.~A.,
        1993.
        Phys. Rev. A 47, R3479-R3482.

\bibitem{Quarles93b}
        Liu, J., Kahler, D.~L., Quarles, C.~A.,
        1993.
        Phys. Rev. A 47, 2819-2826.

\bibitem{Quarles93c}
        Quarles, C.~A., Liu, J.,
        1993.
        Nucl. Instrum. Methods B 79, 142.

%%%%%%%
\bibitem{HipplerSchneider1994}
        Hippler, R., Schneider, H.,
        1994,
        Nucl. Instrum. Methods A 87, 268.

%%%%%%%
\bibitem{Smirnov1977}
        Smirnov, A. I.,
        1977.
        Sov. Phys. -- Nucl. Phys. 25, 548-552.
%%%%%%%
\bibitem{GhilenceaToaderDiaconu1995}
        Ghilencea, D., Toader, O., Diaconu, C.
        1995.
        Romanian Rep. in Phys. 47, 185-196.
%%%%%%%% ++
\bibitem{GavrilaMaquetVeniard1985}
        Gavrila, M., Maquet, A., V\'eniard, V.,
        1985.
        Phys.~Rev. A 32, 2537-2540
        (Erratum:  1986. ibid. 33, 2826).
%%%%%%%%
\bibitem{Korol1993a}
        Korol, A.~V.,
        1993.
        J.~Phys.~B 26, 3137-3145.
%%%%%%%%
\bibitem{AmusiaKorol1993}
         Amusia, M.Ya.,  Korol, A.V.,
        1993,
        Nucl. Instrum. Methods  B  79,  146.

%%%%%%%
\bibitem{KrackeEtAl1993}
        Kracke, G., Alber, G., Briggs, J.S., Maquet, A.,
        1993,
        J.~Phys.~B 26, L561.

%%%%%%%
\bibitem{KrackeEtAl1994}
        Kracke, G., Briggs, J. S., Dubois, A., Maquet, A., V\'eniard, V.,
        1994,
        J. Phys. B 27, 3241.
%%%%%%%%%
\bibitem{DonderaFlorescuPratt1996}
        Dondera, M., Florescu, V., Pratt, R.~H.,
        1996.
        Phys. Rev. A 53, 1492.
%%%%%%%%
\bibitem{Gorshkov1964b}
        Gorshkov, V.~G.,
        1964.
        Zh.~Eksp.~Teor.~Fiz. 47,  1984-1988.
%%%%%%%
\bibitem{FlorescuGavrila1976}
        Florescu, V., Gavrila, M.,
        1976.
        Phys.~Rev.~A 14, 211-235.
%%%%%%%
\bibitem{WongYeh1985}
         Wong, M.~K.~F., Yeh, E.~H.~Y.,
         1985.
         J. Math. Phys. 26, 1701-1710.
%%%%%%%%
\bibitem{ZapryagaevManakovPalchikov1985}
         Zapryagaev, S.~A., Manakov, N.~L., Palchikov, V.~G.,
         1985.
         The Theory of One- and Two-Electron Multicharged Ions.
         Energoatomizdat, Moscow (in Russian).
%%%%%%%
\bibitem{ManakovOvsyannikovRappoport1986}
        Manakov, N.~L., Ovsyannikov, V.~D., Rappoport, L.~P.,
        1986.
        Phys.~Rep. 141, 319.
%%%%%%%%  ++
\bibitem{MuCrasemann1988}
        Mu, X., Crasemann, B.,
        1988.
        Phys.~Rev. A 38, 4585-4596.
%%%%%%%% ++
\bibitem{DonderaMarineskuPratt1990}
        Florescu, V., Marinesku, M., Pratt, R.H.,
        1990.
        Phys.~Rev.~A 42, 3844-3851
        (Erratum:  1991. ibid. 43, 6432).
%%%%%%%%
\bibitem{BergstromSuricPiskPratt1993}
        Bergstrom, P.M. Jr., Suri{\'c}, T., Pisk, K., Pratt, R.H.,
        1993.
        Phys.~Rev.~A 48, 1134-1162.
%%%%%%%
\bibitem{ManakovMaquetMarmoSzymanovski1998}
        Manakov, N.L., Maquet, A., Marmo, S.I., Szymanovski, C.,
        1998.
        Phys.~Lett. 237A, 234-239.
%%%%%%%%
\bibitem{Szmytkowski2002}
        Szmytkowski, R.,
        2002.
        Phys.~Rev.~A 65, 012503.
%%%%%%%%
\bibitem{Yakhontov2003}
        Yakhontov, V.,
        2003.
        Phys. Rev. Lett. 91, 093001.
%%%%%%%%
\bibitem{Szmytkowski2004}
        Szmytkowski, R., Mielewczyk, K.,
        2004.
        J. Phys. B 37, 3961-3972.
%%%%%%% ++
\bibitem{Rosenberg1987_1990_1991}
        Rosenberg, L.,
        1987.
        Phys.~Rev.~A 36, 4284-4289;
        1990. ibid. 42,  5319-5327;
        1991. ibid. 44,  2949-2954.
%%%%%%%
\bibitem{GribakinIvanovKorolKuchiev1999}
        Gribakin, G. F., Ivanov, V. K., Korol, A. V., Kuchiev, M. Yu.,
        1999, J. Phys. B 32, 5463;
        2000, ibid. 33, 821.

%%%%%%%
\bibitem{BMX}
        Varshalovich, D.~A., Moskalev, A.~N., Khersonskii, V.~K.,
        1988.
        Quantum Theory of Angular Momentum. World Scientific, NY.
%%%%%%%%
\bibitem{Geltman1996}
        Geltman, S.,
        1996.
        Phys.~Rev.~A 53, 3473-3483.
%%%%%%%
\bibitem{Sternheimer}
        Sternheimer, R.~M.,
        1954.
        Phys. Rev. 96, 951-968.
%%%%%%%
\bibitem{DalgarnoLewis}
         Dalgarno, A., Lewis, J.T.,
        1955.
        Proc. R. Soc. London A 223, 70-74.

%%%%%%%%%%%%%%%%%%%%%%%%%%%%%%%%%%%%%%%%%%%%%%%%%%%%%%%%%%%%%%%%%%%%%%%%
%%%%%%%%
\bibitem{ElwertHaug69}
        Elwert, G., Haug, E.,
        1969.
        Phys.~Rev. 183, 90-105.

\bibitem{ShafferPratt1997}
        Shaffer, C.D. and Pratt, R.~H.,
        1997.
        Phys.~Rev.~A 56, 3653-3658.

\bibitem{Hostler1964}
        Hostler, L.,
        1964.
        J. Math. Phys. 5, 591.
%%%%%%%
\bibitem{Nordsieck}
        Nordsieck, A.,
        1954.
        Phys.~Rev. 93, 785-787.
%%%%%%%%%%
\bibitem{AvdPratt99}
        Avdonina, N. B., Pratt, R. H.,
        1999,
        J. Phys. B  32, 4261-4276.


\end{thebibliography}
\end{document}